\documentstyle[12pt,fleqn]{article}
\def\singlespace {\smallskipamount=3.75pt plus1pt minus1pt
                  \medskipamount=7.5pt plus2pt minus2pt
                  \bigskipamount=15pt plus4pt minus4pt
                  \normalbaselineskip=15pt plus0pt minus0pt
                  \normallineskip=1pt
                  \normallineskiplimit=0pt
                  \jot=3.75pt
                  {\def\smallskip {\vskip\smallskipamount}}
                  {\def\medskip   {\vskip\medskipamount}}
                  {\def\bigskip   {\vskip\bigskipamount}}
                  {\setbox\strutbox=\hbox{\vrule
                    height10.5pt depth4.5pt width 0pt}}
                  \parskip 7.5pt
                  \normalbaselines}
\def\middlespace {\smallskipamount=5.625pt plus1.5pt minus1.5pt
                  \medskipamount=11.25pt plus3pt minus3pt
                  \bigskipamount=22.5pt plus6pt minus6pt
                  \normalbaselineskip=22.5pt plus0pt minus0pt
                  \normallineskip=1pt
                  \normallineskiplimit=0pt
                  \jot=5.625pt
                  {\def\smallskip {\vskip\smallskipamount}}
                  {\def\medskip   {\vskip\medskipamount}}
                  {\def\bigskip   {\vskip\bigskipamount}}
                  {\setbox\strutbox=\hbox{\vrule
                    height15.75pt depth6.75pt width 0pt}}
                  \parskip 11.25pt
                  \normalbaselines}
\def\doublespace {\smallskipamount=7.5pt plus2pt minus2pt
                  \medskipamount=15pt plus4pt minus4pt
                  \bigskipamount=30pt plus8pt minus8pt
                  \normalbaselineskip=30pt plus0pt minus0pt
                  \normallineskip=2pt
                  \normallineskiplimit=0pt
                  \jot=7.5pt
                  {\def\smallskip {\vskip\smallskipamount}}
                  {\def\medskip   {\vskip\medskipamount}}
                  {\def\bigskip   {\vskip\bigskipamount}}
                  {\setbox\strutbox=\hbox{\vrule
                    height21.0pt depth9.0pt width 0pt}}
                  \parskip 15.0pt
                  \normalbaselines}

\include{dspace12}
\def\be{\begin{equation}}
\def\ee{\end{equation}}
\def\bea{\begin{eqnarray}}
\def\eea{\end{eqnarray}}
\def\nn{\nonumber}
\def\th{\theta}
\def\ph{\phi}
\def\lt{\left}
\def\rt{\right}
\def\sect #1{\setcounter{equation}{0}}

\begin{document}
%\singlespace
\middlespace
%\doublespace
%%%%%%%%%%%%%%%%%%%%%%%%%%%%%%%%%%%%%%%%%%%%%%%%%%%%%%%%%%%%%%%%%%
\begin{center}
{\LARGE {A  conformal scalar dyon black hole solution}}
\end{center}
\vspace{1.5in}
%\vspace{12pt}
\begin{center}
{\large{K. S. Virbhadra\footnote[1]{Present address\ :\ Departamento de
F\'{\i}sica Te\'{o}rica, Universidad del Pa\'{i}s Vasco, 48080 Bilbao, Spain,
E-mail\ :\ WTXVIXXK@LG.EHU.ES} \
\  and J. C. Parikh \\
Physical Research Laboratory \\
Navrangpura, Ahmedabad - 380 009, India}}
\end{center}
\vspace{1.0in}
\begin{abstract}
An exact solution of Einstein - Maxwell - conformal scalar field
equations is given, which is a black hole solution and has three
parameters: scalar charge, electric charge, and magnetic charge.
Switching off the magnetic charge parameter yields the solution
given by Bekenstein. In addition the energy of the conformal scalar
dyon black hole is obtained.
\end{abstract}
\vspace{1.0in}
\begin{center}
{\bf To appear in Physics Letters B}
\end{center}
\middlespace
\newpage
It is known that Maxwell's equations are conformally invariant
in four dimensions, whereas the ordinary massless scalar equation
is conformally invariant  in two dimensions. However, there
are scalar equations, which are conformally invariant in any
arbitrary finite dimension (dimension greater than one) $[1]$.
For instance, in four
dimensions one has the following conformally invariant scalar equation :
\be
\Psi_{,i}^{\ \ ;i} - \frac{1}{6} R \Psi = 0
\ee
$\Psi$ is the conformal scalar field and $R$ is the scalar curvature.
Comma and semicolon denote, respectively, the partial and
covariant derivatives. We use the geometrized units $(G = 1, c =
1)$ and follow the convention that the Latin indices take values 0 to 3
and Greek indices take values 1 to 3. $x^0$ is the time coordinate.
The energy-momentum tensor for the conformal scalar field given
by Eq. $(1)$ differs from that
of the ordinary massless scalar field and it has several interesting properties
$[2-4]$. Inspired by this,
Bekenstein $[4]$ obtained a static and spherically symmetric exact
solution of Einstein-Maxwell-conformal scalar field equations
which is characterized by two parameters: scalar
charge and electric charge.
The solution obtained by Bekenstein is asymptotically flat and
has an event horizon, but the scalar field diverges on the event
horizon. Therefore, he suggested that the solution cannot be
interpreted as a black hole solution. However, his subsequent
analysis $[5]$ revealed that, the infinity in the conformal scalar field is
not associated with an infinitely high potential barrier, no test
particle trajectories terminate at horizon at finite proper time,
and tidal accelarations  remain bounded at the horizon. He
concluded that the horizon is physically regular and
therefore the solution obtained by him is a black hole solution.
It was conjectured that the stationary black holes could be
parametrized only by mass, angular momentum, electric and
magnetic charges and this was paraphrased by Wheeler as ''\,black
holes have no hair\,''. Having shown that  the solution obtained by him
is a black hole solution, Bekenstein stressed that
the scalar charge should also be  included as a black hole parameter.

Although the existence of magnetic monopoles
is not yet confirmed, it has drawn attention of many
theoreticians (see in reference $6$).
As Bekenstein's solution  is not enriched by the
magnetic charge parameter, it is of interest to obtain
a conformal scalar dyon (CSD) black hole solution (of
Einstein-Maxwell-conformal scalar field
equations) which is characterized by scalar, electric
and magnetic charges.

The Einstein-Maxwell-conformal field equations $[4]$ are
\be
R_{ij} - \frac{1}{2} g_{ij} R = 8 \pi \Sigma_{ij}
\ee
where
%%%%%%%%%%%%%%%%%%%%%%%
\be
\Sigma_{ij} = \lt(\Theta_{ij} + E_{ij}\rt) \lt(1 - \zeta^2 \Psi^2\rt)^{-1}
\ee
%%%%%%%%%%%%%%%%%%%%%%%
\be
\Theta_{ij} =  S_{ij} - \frac{1}{6} {\lt(\Psi^2\rt)}_{,i;j}
                     + \frac{1}{6} {\lt(\Psi^2\rt)}_{,l;m}
                         g^{lm} g_{ij}
\ee
%%%%%%%%%%%%%%%%%%%%%%
\be
S_{ij} = \Psi_{,i} \Psi_{,j} - \frac{1}{2} g_{ij}\ g^{kl}
                                     \ \Psi_{,k} \Psi_{,l}
\ee
%%%%%%%%%%%%%%%%%%
\be
E_{ij} = \frac{1}{4\pi} \lt[ F_{ik} F_{jl} - \frac{1}{4} g_{ij}
                                      \  g^{mn} F_{mk} F_{nl} \rt] g^{kl}
\ee
($S_{ij}$  and $E_{ij}$, respectively, are
the energy-momentum tensors of the ordinary scalar field and
electromagnetic field)
%%%%%%%%%%%%%%%%%%%%%
\be
\zeta = \sqrt{\frac{4 \pi}{3}}
\ee
%%%%%%%%%%%%%%%%%%%%%%%%%%%%%%%
\be
\lt(\sqrt{-g}\ F^{ij}\rt)_{,j} = 0
\ee
%%%%%%%%%%%%%%%%%%%%%%%%%%%%%%%%%%%
\be
\lt(\sqrt{-g}\ \  {{}^{\star}F}^{ij}\rt)_{,j} = 0
\ee
%%%%%%%%%%%%%%%%%%%%%%%%%%%%%%%%%%
${{}^{\star}F}^{ij}$, the dual of the electromagnetic
field tensor $F^{ij}$, is
\be
{{}^{\star}F}^{ij} = \frac{1}{2 \sqrt{-g}} \varepsilon^{ijkl} \ F_{kl}
\ee
($\varepsilon^{ijkl}$ is the Levi-Civita tensor density). Recall
that the conformal scalar equation is given by Eq. $(1)$.

An exact solution of the above equations, characterized by the
scalar charge, electric charge, and magnetic charge is given by
the line element,
%%%%%%%%%%%%%%%%%%
\be
ds^2 = \lt(1 - \frac{Q_{CSD}}{r} \rt)^2\ dt^2 - \lt(1 -
            \frac{Q_{CSD}}{r}\rt)^{-2} dr^2 - r^2 \lt(d\th^2 +
                                    sin^2\th\ d\ph^2\rt)
\ee
%%%%%%%%%%%%%%%%
the conformal scalar field,
\be
\Psi = \sqrt{\frac{3}{4 \pi}}
       \lt( \frac{q_s}{r-Q_{CSD}} \rt)
\ee
where
\be
Q_{CSD}\ = \ \sqrt{{q_s}^2 + {q_e}^2 + {q_m}^2}
\ee
%%%%%%%%%%%%%%
and the non-vanishing components of the electromagnetic field tensor,
\be
F_{rt} = \frac{q_e}{r^2}
\ee
%%%%%%%%%%%%%%%
\be
F_{\th \ph} = q_m sin\th
\ee
%%%%%%%%%%%%%%%%%%%%%
where $q_s, q_e,$ and $q_m$, respectively, denote the scalar
charge, electric charge, and magnetic charge. For convenience we
call $Q_{CSD}$  {\it conformal scalar dyon  charge}, which is the square root
of the sum of the squares of the scalar, electric, and magnetic charges.
The solution given by us has an event horizon at $r = Q_{CSD}$.

For the CSD black hole solution, given
by $(11) - (15)$, the non-vanishing components of $\Theta^i_k$ and
$E^i_k$ are
%%%%%%%%%%%%%%%%%%%%%%%%%%%%%%%%%%%%%%%%%%%%%%%%%%%%%
\be
\Theta^0_0 = \Theta^1_1 = - \Theta^2_2 = - \Theta^3_3 =
           \frac{{q_s}^2 \lt( r - 2 Q_{CSD} \rt)}
                {8 \pi r^3 \lt( r -  Q_{CSD} \rt)^2}
\ee
%%%%%%%%%%%%%%%%%%%%%%%%%%%%%%%%%%%%%%%%%%%%%%%%%%%
\be
E^0_0 = E^1_1 = - E^2_2 = - E^3_3 = \frac{{q_e}^2 + {q_m}^2}
                                         {8 \pi r^4}
\ee
%%%%%%%%%%%%%%%%%%%%%%%%%%%%%%%%%%%%%%%%%%%%%%%%
and the non-vanishing components of the Einstein tensor $G^i_j
\equiv R^i_j - \frac{1}{2} g^i_j R $ and $\Sigma^i_j$ are
\be
G^0_0 = G^1_1 = -G^2_2 = -G^3_3 = 8 \pi \Sigma^0_0 = 8 \pi \Sigma^1_1
= -  8 \pi \Sigma^2_2 = - 8 \pi \Sigma^3_3 = \frac{Q_{CSD}^2}{r^4}
\ee
%%%%%%%%%%%%%%%%%%%%%%%%%%%%%%%%%%%%%%

It is of interest to obtain the gravitational energy of the
CSD black hole. There are many
prescriptions for obtaining energy and momentum in curved
spacetimes $[7]$.
Among them  Weinberg's prescription is one of the most handy.
The expression for energy in Weinberg's prescription is $[8]$
\be
E = \frac{1}{16 \pi} \int \int \lt(
                               \frac{\partial h^{\alpha}_{\alpha}}
                                    {\partial x_{\beta} }
                              - \frac{\partial h^{\alpha \beta}}
                                     {\partial x^{\alpha}}
                           \rt) n_{\beta} r^2 sin\th d\th d\ph
\ee
where $n_1 = x/r,\ n_2 = y/r,\ n_3 = z/r$,\ $r^2 = x^2 +y^2 +z^2$
and
\be
h_{ij} = g_{ij} - \eta_{ij}
\ee
$\eta_{ij}$ is the Minkowski metric and indices on $h_{ij}$ or $\frac{\partial}
{\partial x^i}$ are raised or lowered with help of $\eta 's$. It
is known that  Weinberg's prescription for obtaining energy
of a general relativistic system gives correct result if
calculations are carried out in coordinates in which the metric
$g_{ij}$ approaches the Minkowski metric $\eta_{ij}$ at great
distances from the system under study. Therefore, one writes the
metric $(11)$ in  Kerr-Schild Cartesian  coordinates $(T,x,y,z)$
$[9]$.
\be
ds^2 = dT^2 - dx^2 - dy^2 - dz^2 - \frac{2 Q_{CSD}
\lt(1-\frac{Q_{CSD}}{2r}\rt)}
                                        {r}
      \lt[dT+\frac{x dx + y dy +zdz}{r}\rt]^2
\ee
The coordinates $t,r,\th,\ph$ in $(11)$ and $T,x,y,z$ in $(21)$
are related through
\bea
T = t - r + \int \lt( 1 - \frac{Q_{CSD}}{r} \rt)^{-2} \ dr \nn\\
x = r\  sin\th \ cos\ph \nn\\
y = r\  sin\th \ sin\ph \nn\\
z = r\  cos\th \nn\\
\eea
With  $(20)$ and $(21)$, one has
\be
h_{\alpha \beta} = \frac{ - 2 Q_{CSD} \lt(1 -  \frac{Q_{CSD}}{2r}\rt)}{r^3}
                              \ x^{\alpha} x^{\beta}
\ee
Using $(23)$ in $(19)$, a straightforward calculation yields
\be
E\ = \ Q_{CSD} \lt( 1 - \frac{Q_{CSD}}{2 r} \rt)
\ee
Therefore, the total gravitational energy of a CSD black hole (
$r$ approaching infinity in Eq. $(24)$ ) is given by its CSD charge.

We have given an exact solution of CSD black hole which has
three parameters: scalar, electric, and magnetic charges. This
solution is the magnetic charge generalization of the Bekenstein
solution. It is obvious that the metric of the CSD black hole is
the same as the well known Reissner-Nordstr\"{o}m  (R-N) metric when mass
and charge parameters both are set to the CSD charge. Like the R-N
geometry this has a curvature singularity at $r = 0$. However,
it differs from the R-N metric as it has only one  horizon
($r = Q_{CSD}$). The total gravitational energy of the CSD black
hole is given by its CSD charge, whereas for the R-N black hole the
total energy is given by its mass parameter $[10]$. For the
charge parameter greater than the mass parameter for the R-N geometry,
there is no event horizon and the singularity is naked. However,
for the CSD geometry, the singularity is always covered by its
event horizon.

In passing we remark that Bekenstein as well
as we have  not considered the self interaction terms in the conformal scalar
field equation and in  the corresponding
energy-momentum  tensor. It is of interest to include them
and obtain solutions for CSD black hole. Further, it is
essential to study the stability of the CSD black hole.

\begin{flushleft}
{\bf References}

$[1]$ R. M. Wald, {\it General Relativity (The University of
Chicago Press, 1984)} p447\\
$[2]$ C. G. Callan, Jr., S. Coleman, and R. Jackiw, Ann. Phys.
(N.Y.) {\bf 59}  \ (1970) \ 42\\
$[3]$ L. Parker, Phys. Rev. D {\bf 7}\ (1973)\ 976\\
$[4]$ J. D. Bekenstein, Ann. Phys. (N.Y.) {\bf 82} \ (1974)\ 535\\
$[5]$ J. D. Bekenstein, Ann. Phys. (N.Y.) {\bf 91}\ (1975) \ 75\\
$[6]$ A. S. Goldhaber and W. P. Trower, Am. J. Phys. {\bf 58}
\ (1990)\ 429\\
$[7]$ J. D. Brown and J. W. York, Jr, Phys. Rev. D  {\bf 47}\ (1993)
\ 1407\\
$[8]$ S. Weinberg, {\it Gravitation and Cosmology,
Principles and applications of the general theory of relativity,
(John Wiley and Sons. NY 1972)} p165\\

$[9]$ B. Carter, Phys. Rev. {\bf 174} \ (1968)\ 1559\\
$[10]$ K. P. Tod, Proc. R. Soc. Lond. A {\bf 388} \ (1983)\ 467

\end{flushleft}
\end{document}